\documentclass[aps,pre,twocolumn,showpacs,superscriptaddress]{revtex4}
\usepackage{graphicx}
\usepackage{amsmath}
\usepackage{setspace}
\usepackage{dcolumn}
\usepackage{bm}
\usepackage{epstopdf}

\begin{document}
\title{Strong Frequency Dependence in Over-damped Systems}
\author{Ay\c{s}e Ferhan Ye\c{s}il}
 \email{yesil@bilkent.edu.tr}
\author{M.~Cemal Yalabik}
\affiliation{Department of Physics, Bilkent University, 06800
Ankara, Turkey}
\date{\today}
\begin{abstract}
Strong frequency dependence is unlikely in diffusive or over-damped systems. When exceptions do occur, such as in the case of stochastic resonance, it signals an interesting underlying phenomenon. We find that such a case appears in the motion of a particle in a diffusive
environment under the effect of periodically oscillating retarded force emanating from the boundaries. The amplitude for the expectation value of position has an oscillating frequency dependence, quite unlike a typical resonance. We first present an analysis of the associated Fokker-Planck equation, then report the results of a Monte Carlo simulation of the
effect of a periodic perturbation on a totally asymmetric simple exclusion process (TASEP) model with single species. This model is known to exhibit a randomly moving shock profile, dynamics of which is a discrete realization of the Fokker-Planck equation. Comparison of relevant quantities from the two analyses indicate that the same phenomenon is apparent in both systems.

\end{abstract}
\maketitle
\section{Introduction}
Diffusive systems cannot support resonances in the usual sense, a periodic force cannot build oscillations that grow with each cycle. The frequency response of systems vary monotonously. We present an over-damped system which has an oscillating frequency response due to the presence of a position dependent effective force. Such conditions arise for objects under the influence of retarded effects from boundaries an example of which we also provide.

The Fokker-Planck (FP) equation \cite{Fokker,Planck,Kolmogorov,Risken,Hanggi} describes the time dependence of the probability distribution $P(x,t)$ of the 
position of a particle in a diffusive environment: 
\begin{equation}
 \frac{\partial P}{\partial t} = \frac{\partial}{\partial x} \left [
-\gamma P (-\frac{\partial V}{\partial x} +F_{ret}) +D\frac{\partial P}{\partial x} \right ]
\label{FP}
\end{equation}
where $\gamma$ is the drift constant relating velocity to force and $D$ is the coefficient of the 
diffusion term. The potential energy $V$ is related to the time-independent force field the particle is in,
while $F_{ret}$ represents a time and position dependent driving force which will
be relevant to our problem. Specifically, we will consider a force which emanates from sinusoidally varying
boundary effects, propagating with a constant velocity into the interval of interest, so that the space
dependence of the force is also sinusoidal.

Considerable amount of work has been carried out on the behaviour of $P$, under the 
effect of various types of potentials. Analytical solutions are usually restricted to simpler forms of potentials.
The equation may be treated as a Schr\"{o}dinger equation with imaginary time, and the time dependent solutions
have a relaxational type of behaviour. This is the consequence of the assumption of absence of any 
inertia (or ``memory'') in the system: The time rate of change of $P$ is described in terms of the value of $P$
at time $t$. However, we find ``resonance" behavior, in a looser context, as a consequence of Eqn.~\ref{FP}, which corresponds to amplified response at a sequence of frequencies, based on
matching of the waveform of the driving force to the interval between the boundaries.

A number of other stochastic systems with damped dynamics do display
non-monotonous frequency response to sinusoidal drives. In Brownian
motors, the particle moves in a (usually ratchet-type) static potential
and the sinusoidal force is used to drive the particle over the ``easier"
barrier \cite{bartussek}. (The asymmetry in the direction may be
achieved either by the form of the static potential or the form of the
time-dependent forces.) Alternatively, a high-frequency drive may
effectively act to average over portions the static potential, on a length
scale which depends on the properties of the drive. This results in a
``vibrational resonance" \cite{vibres} with transport properties strongly
dependent on parameters associated with the drive. The force may take the
form of a superposition of a number of sinusoidal (or rectangular)
functions, in which case the non-linear effects may lead to interesting
effects at certain combination of frequencies \cite{ratchett}. Besides the
models we mention here, the reader is referred to a review by H{\"a}nggi
and Marchesoni \cite{vibres2} for a detailed discussion of these and other similar
systems.

Sinusoidal drive on a damped system has also been studied in relation to ``stochastic resonance" \cite{StoRes,Jung,Gammaitoni1,Jung2,stocres,Gammaitoni2}. 
This relates basically to the motion of a damped particle in a double well potential on which a random diffusive force as well as the sinusoidal force is acting. 
It had been demonstrated that there would be an optimal magnitude for the random force which amplifies transitions between the potential minima in 
synchronism with the periodic drive. Jung and H\"{a}nggi \cite{Jung} solved the corresponding FP equation, demonstrating that the time autocorrelation function of a periodically driven bistable over-damped system sustains undamped oscillations. The term ``resonance'' here is associated with amplified response at a certain value of the magnitude of the random force. The process has been used to model a very wide spectrum of phenomena, ranging for example from the effect of seasonal changes on population systems \cite{Bio}
to the dependence of stock prices on periodic information flow \cite{Li}, as well as a multitude of biological phenomena\cite{Ushakov, Martignoli}.

All of the above processes incorporate time-independent multiple potential wells and position independent periodic forces. In contrast to, the phenomena we report is associated with a static potential confined to the boundaries and a periodic retarded force also emanates from the boundaries. The effects of a force field propagating with a finite velocity 
(which in our case is equivalent to a position dependent time-periodic force) has not received much attention. 
Relevant work include the analysis of a diffusive system which injects particles (which are annihilated upon contact with one another) 
at a boundary \cite{Zygouras}, and the derivation of an effective force in a Hamiltonian system with a position-dependent sinusoidal force coupled to a 
bath of oscillators (resulting in the random force) \cite{Shit}.

It may seem contradictory to discuss the presence of propagating effects within the context of a memoryless, over-damped system. However, the underlying mechanism which drives the diffusion dynamics can be distinct from the process which generates the effective force. This is the case, for example, in Totally Asymmetric Simple Exclusion Process (TASEP) systems where the diffusion of a persistent shock front in the system \cite{asolversguide} is influenced by the boundary conditions after a delay. Thus, TASEP forms a discrete realization of the aforementioned FP system. We first analyse the FP system (Sec. II/A), we then report the results of a Monte Carlo study carried out on the TASEP system as well  (Sec. II/B).

\section{Analysis and Discussions}
\subsection{Fokker-Planck Equation}
We will first describe our analysis of the FP equation.
In particular, we take the time-independent potential $V(x)$ to be given by
\begin{equation}
V(x) = \left \{ \begin{array}{ll}0 & \mbox{if }|x|< \frac{L}{2} \\ 
   V_0 (|x|-L/2)^2/x_0^2 & \mbox{otherwise.} \end{array} \right . \label{pot}
\end{equation}
The quadratic structure softens the boundaries at $\pm L/2$ with a range related to $x_0$.
We also assume that a time dependent force acts on the particle, synchronously from the two boundaries,
but retarded in time (with a propagation speed $v$) in proportion to the distance to the boundaries:
\begin{eqnarray}
 F_{ret}(x,t) &=& F_0 \sin \left [ \omega\left (t-\frac{L/2+x}{v}\right )\right ] \nonumber \\
              & & +  F_0 \sin \left [ \omega\left (t-\frac{L/2-x}{v}\right )\right ] \nonumber \\
 &=& 2F_0 \cos \frac{\omega x}{v} \sin \left [\omega\left (t-\frac{L}{2v}\right )\right ]. \label{force}
\end{eqnarray}
It can be observed that the retarded force results in a position dependent amplitude in the oscillation. This amplitude is not frequency dependent, however we will show that the magnitude of the response to this force does depend on how the wavelength compares to the size of the system.
We write Eqn.~\ref{FP} in scaled form
\begin{eqnarray}
 \frac{\partial P(z,\theta)}{\partial \theta} &=& \Gamma \frac{\partial}{\partial z} \left [
P \; \frac{d \widetilde{V}(z)}{d z} -\epsilon \; \cos(2\pi z/\lambda)\sin (\theta) \right ]\nonumber \\
 &+& \overline D \frac{\partial^2 P}{\partial z^2} \label{scaled_FPE}
\end{eqnarray}
where we have used the dimensionless quantities in Table \ref{table:tablo} with unitless potential:
\begin{eqnarray}
\widetilde{V}(z) &=& \left \{ \begin{array}{ll}0 & \mbox{if }|z|< \frac{1}{2} \\ 
    (|z|-1/2)^2L^2/x_0^2 & \mbox{otherwise.} \end{array} \right .  \nonumber 
\end{eqnarray}

\begin{table}[width=8.6cm]
\begin{center}
\begin{tabular}{    l  l    }
\hline
   \multicolumn{2}{ c }{Dimensionless Quantities} \\
   \hline 
   \\
   $\theta  = \omega (t-L/(2v) )$   &    $ \Gamma = \gamma V_0/(\omega L^2)$  \\
         $z = x/L$                  &    $\epsilon = 2F_0L/V_0 $ \\
   $\lambda = 2\pi v/(\omega L)$    &    $\overline D = D/(\omega L^2) $\\ [0.25cm]
\hline
\hline
 \end{tabular}
\end{center}
\caption{Dimensionless quantities that are used in scaling the Fokker-Planck equation.}
\label{table:tablo} 
\end{table}

The parameter $\lambda$, besides representing the wavelength of the time dependent
force relative to the distance variable $z$, is also proportional to the period of oscillation:
$\tau = 2\pi/\omega = \lambda L/v$.

We have solved equation \ref{scaled_FPE} numerically for various values of the parameters. The probability
density $P(z,\theta)$ was solved on a mesh of 256 points in the $z$ direction, 
corresponding to the $x$-coordinate values for $|x|<L/2$ plus the two boundary regions 
which were taken to be $3x_0$ wide each. The equation was integrated numerically in the
time variable $\theta$ with step sizes such that $\Delta \theta/(\Delta z)^2 \leq 0.1$. The integration
for the period of $\theta=2\pi$ was repeated 10 times, which was sufficient for convergence,
{\em i.e.} for obtaining no appreciable change in $P(z,\theta)$.

We then calculate the expectation value of position as a function of $\theta$:
\begin{equation}
\overline z(\theta) = \int_{-\infty}^{\infty} dz \;z\; P(z,\theta). \nonumber 
\end{equation}
The size of this oscillation is parametrized though the fundamental Fourier coefficients:
\begin{eqnarray} 
C &=& \frac{1}{2\pi}\int_0^{2\pi} \overline z(\theta) \cos(\theta) d \theta \nonumber \\
S &=& \frac{1}{2\pi}\int_0^{2\pi} \overline z(\theta) \sin(\theta) d \theta . \label{sincos}
\end{eqnarray}
\begin{figure}       
  \begin{center}
     \includegraphics[width=8.6cm]{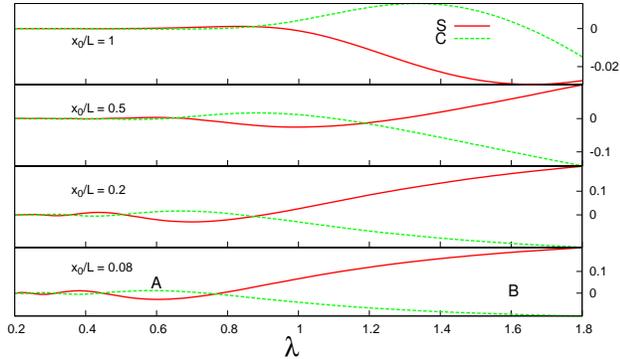}
   \end{center}
  \caption{\label{fig:coklu_plot} The magnitude of the oscillation of the position expectation value as a function of wavelength $\lambda$
  for various values of $x_0/L$, the boundary smoothness. (Other parameters which enter Eqn.~\ref{scaled_FPE} are $\Gamma=1,~\overline D=0.225$ and $\epsilon = 1$.) 
  The continuous (dotted) line represents the in phase (out of phase) fundamental component $S$ ($C$) in Eqn.~\ref{sincos} of the oscillation.
  For large values $\lambda$, the value of $C$ saturates due to the system width. Note the different scales for the response in the figures.}
\end{figure}

Fig.~\ref{fig:coklu_plot} displays that the response of the system has strong frequency dependence. The amplitude of the oscillations grow as $x_0/L$ decreases, when the boundaries become sharper.
A wider boundary allows more oscillations (as a function of position)
in the probability density, diminishing the variations in $\overline z$. As the boundary region expands, so do the
features on the plot, implying longer wavelengths. For values of $\lambda$ much larger than one, the response monotonically increases 
to its asymptotic value. Note also that the ``in phase'' component $S$ dominates the response for smaller $x_0/L$. The size of $\epsilon$ was chosen to obtain a response magnitude comparable to that obtained from the Monte Carlo analysis.

\begin{figure}       
  \begin{center}
     \includegraphics[width=8.6cm]{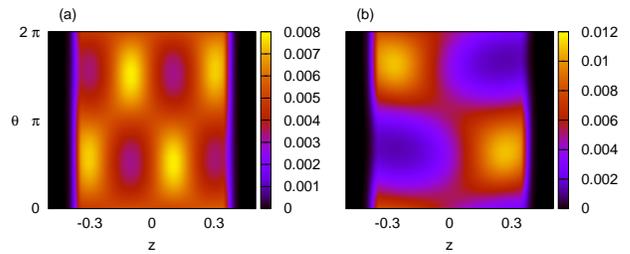}
   \end{center}
  \caption{\label{fig:ikili_fp} The probability densities of marked points in Eqn.~\ref{scaled_FPE}. (a) Corresponds to point {\bf A} ($\lambda = 0.6$), consecutive dark and light patterns along the x-axis, indicate standing waves of two wavelengths that fits to the lattice size. (b) Corresponds to point {\bf B} ($\lambda = 1.6$), it can be seen that only one wavelength is supported by the system. } 
\end{figure}

The extrema of the response correspond to matching of the wavelength of the driving force to the effective length
of the diffusion region $L$, plus the boundary region. Fig.~\ref{fig:ikili_fp} displays the probability densities
for points {\bf A} and {\bf B} in Fig.~\ref{fig:coklu_plot}. Note that for $\theta\sim \pi /4$ and $z\sim 0$, the probability densities increase with $z$ in both 
plots $(a)$ and $(b)$, consistent with a sinusoidal drive. However, when an even (odd) number of maxima are present, the position expectation value for these cases become
negative (positive), producing oscillations as a function of wavelength. As more and more wavelengths ``fit'' into the system (smaller $\lambda$), the change in the expectation value of the particle position
becomes less and less discernible. 
\subsection{TASEP}
We now turn to the TASEP system which provides a discrete realization of this phenomenon on a one dimensional lattice. Its dynamics is described by the following transition probabilities in time $dt$ : On the leftmost site $0 \to 1$ with probability $\alpha dt$, inside the bulk $01 \to 10$ with probability $dt$, and on the rightmost site $1 \to 0$ with probability $\beta dt$. These systems have been studied extensively for time-independent boundary rates. Recently, the effects
of time dependence has also received interest: Popkov {\em et al.} studied a vehicular traffic on highways under periodically changing green and red-lights \cite{schutz} and Basu {\em et al.} also showed a frequency dependent modality on similar transport systems \cite{basu}.  

\begin{figure}       
  \begin{center}
     \includegraphics[width=8.6cm]{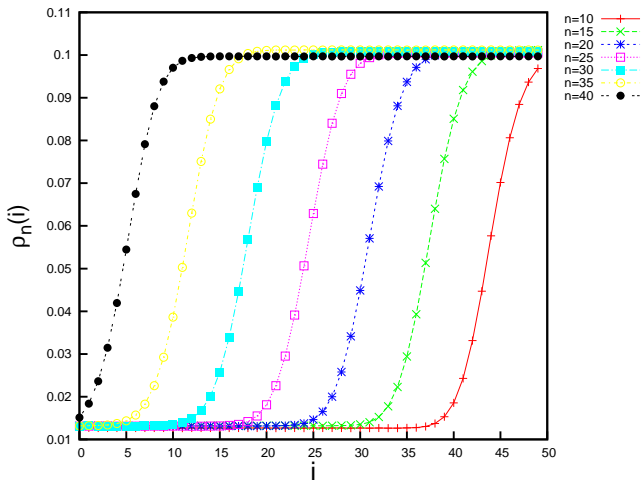}
   \end{center}
  \caption{\label{fig:shockdist} Shock profile distributions for various numbers of particle number. } 
\end{figure}

\begin{figure}       
  \begin{center}
     \includegraphics[width=8.6cm]{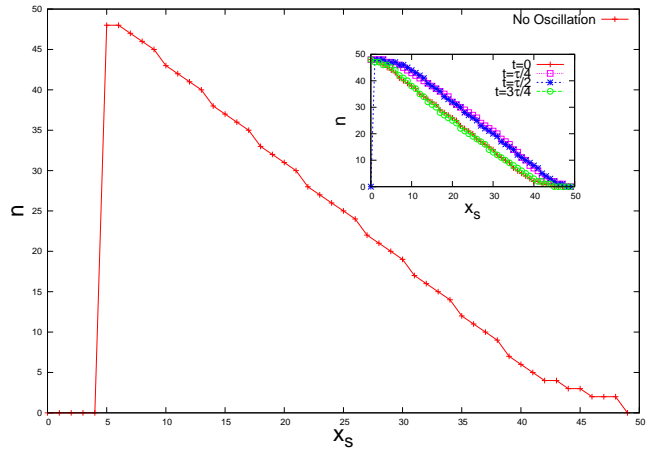}
   \end{center}
  \caption{\label{fig:lineer} The linear relationships between particle number $n$ and shock position $x_s$ for the profiles in Fig.~\ref{fig:shockdist}.  The shock position $x_s$ is defined as the lattice position at which the
density $\rho_n(i)$ has the value
$(\rho_{\mbox{max}} + \rho_{\mbox{min}})/2$ where
$\rho_{\mbox{max}}$ and $\rho_{\mbox{min}}$ are respectively
the maximum and
and minimum values of $\rho_n(i)$. The inset displays the relationship when boundary conditions change sinusoidally with period $\tau$ as discussed in the text. The plots are for four different time values as $t=0$, $t=\tau /4$, $t=\tau /2 $ and $t=3\tau/4$. Here $N=50$ and $\tau = 120$. It can be seen that the linearity of the relationship does not
change appreciably with time, resulting in an approximate relation
of the form $x_s + n \sim N$. Since $x_s$ is determined by interpolation, it in general is not
an integer.
}
\end{figure}

The time-independent system can be solved exactly \cite{exact1, exact2, exact3}. A first order phase transition line separates the high and low density phases for $\alpha= \beta < 1/2$. Along this line the density may be shown to correspond to a superposition of shock profiles which move from one end of the system to the other. These structures are not quantities that are measurable at any instant of time: Occupation statistics of states with a specific number of $n$ particles lead to a shock profile associated with that $n$. Fig.~\ref{fig:shockdist} shows these profiles for a system with 50 sites, for various values of $n$. The linear dependence of the position of the profile on particle number is apparent. Fig.~\ref{fig:lineer} displays this dependence. 

\begin{figure}       
  \begin{center}
     \includegraphics[width=8.6cm]{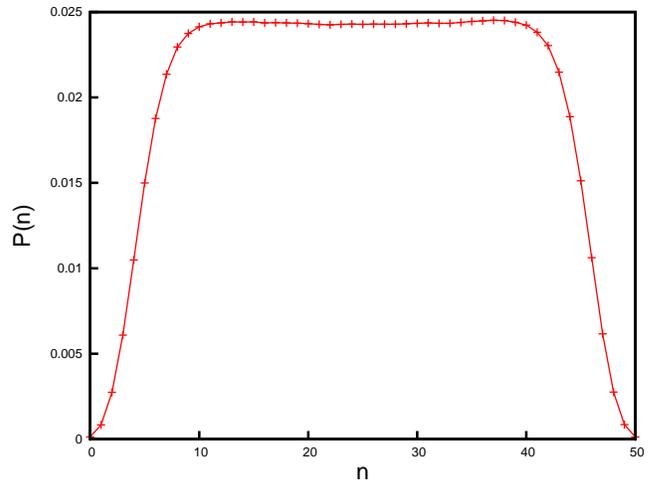}
   \end{center}
  \caption{\label{fig:p_n_dist} Probability distribution $P(n)$ of particle number $n$., for a lattice size of $N=50$ and $\alpha=\beta=0.1$.} 
\end{figure}

A perturbation to the boundary conditions leads to a change in the number of particles in the system. This then acts as an effective force which results in the change of the shock position after a delay. Moreover for small values of $\alpha=\beta$ these shock profiles tend to be evenly distributed within the lattice. In other words, the shock front carries out a random walk in the lattice \cite{asolversguide}. Fig.~\ref{fig:p_n_dist} displays the probability distribution $P(n)$ for the existence of $n$  particles (equivalently contribution from a shock profile at a corresponding position) in the time independent system. This distribution indeed may be interpreted as  the result of a random walk of the profile in a potential of the form $V(x)$. This random walk is ordinarily constrained by the time-independent boundaries: If the shock wave moves too close to the left, increased density at that boundary reduces the particle entry rate, effectively ``pushing'' the profile to the right. A similar mechanism on the exit boundary constrains the profile to the central region of the lattice. This constraint has been represented by a ``free-energy functional" \cite{Arndt} which is quite similar in form to $V(x)$.


\begin{figure}       
  \begin{center}
     \includegraphics[width=8.6cm]{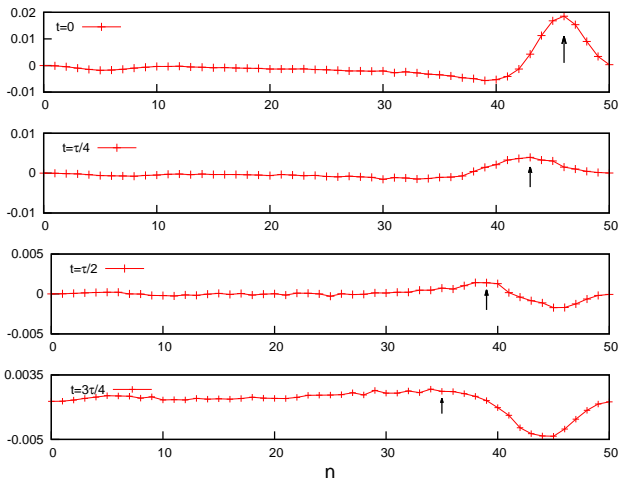}
   \end{center}
  \caption{\label{fig:bump_response} Response of the probability distribution function $P(n)$ to a pulsed $\alpha$ rate. Time averaged values have been subtracted to better display the response. Pulse's magnitude is $\alpha=5$ , other parameters are the same as those in Fig.~\ref{fig:p_n_dist}. The arrows indicate the maxima of the curves to better display the propogation.} 
\end{figure}

The change in the probability distribution in Fig.~\ref{fig:p_n_dist} in response to a pulsed change in the particle entry rate $\alpha$ is shown in Fig.~\ref{fig:bump_response}. The disturbance caused by an influx of extra particles for a short period of time travels along the lattice, with a damped response. (Note the oscillatory nature of the response indicating a travelling wave mechanism as well as damping.) While this motion along the lattice justifies the association of this system with the ``retarded force'' mechanism of Eqn.~\ref{FP}, the relatively strong damping leads to results not as ``clean'' as the ones given in Fig.\ref{fig:ikili_fp} for the FP equation. 

We now proceed with our discussion of the details of our computations of the sinusoidally varying drive.

The MC procedure we used is known as the ``kinetic Monte Carlo" method \cite{kMC}, in which events occur at time intervals $\Delta t = -\ln r / \Omega$ where $r$ is a random number uniformly distributed between $0$ and $1$, and $\Omega$ is the sum of rates of all possible transitions in the system. The event which does take place is also selected randomly with a probability proportional to its rate. Transition rates at the boundaries are assumed to be constant within the duration $\Delta t$, since $\Delta t \ll 2\pi/ \omega$. The Mersenne Twister algorithm was used for generating $r$ \cite{Mersenne1,Mersenne2}. Time dependent probability density calculations are carried out over $10^6$ Monte Carlo steps (MCS). For a lattice of size $N$, we define a MCS as $N^2$ changes in the system. 

We have observed that the time-dependent boundary effects decay appreciably inside the lattice. A small lattice is therefore necessary for boundary effects to be seen inside the bulk. Moreover, broadest range of a random walk  is known to take place for smaller values of $\alpha$ and $\beta$. In order to extend the range of the random walk to the full range between the boundaries, $\alpha=\beta =0.1$ is taken for an anchor point on a lattice of size $N=50$. For the effect to be prominent, boundary rates were varied with a significant amplitude such as $\alpha= 0.1+ 0.099~\mbox{sin}(\omega t)$ and $\beta= 0.1-0.099~\mbox{sin}(\omega t)$, comparable to the average values of $\alpha$ and $\beta$. 

\begin{figure}       
  \begin{center}
     \includegraphics[width=8.6cm]{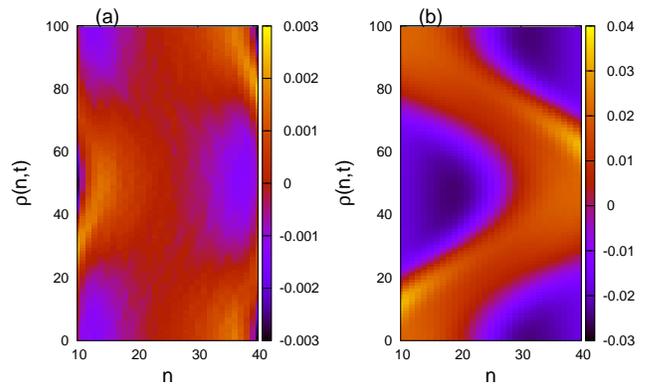}
   \end{center}
  \caption{\label{ikili_mc} Change in probability density $\rho (n,t)$ from its time average for $N=50$ and different periods calculated over $10^6$ MCS. (a) When $\tau=140$, similar to Fig.~\ref{fig:ikili_fp} a, two wavelengths are supported by the system. (b) When $\tau=700$, similar to the results of Fig.~\ref{fig:ikili_fp} b, only one wavelength is supported by the system. In both plots, two extremal coordinates in both sides were excluded to avoid artifacts due to boundary effects. Color scales are arbitrary.} 
\end{figure}

Note that our time-dependent perturbation drives the system between two values of the boundary parameters corresponding to low and high density phases. This drive should then result in an oscillation of the number of particles within the system, hence of the probability density of the shock position. In Fig.~\ref{ikili_mc}, the probability distribution for finding $n$ particles in the system (which is linearly related to shock position as can be seen in Fig.~\ref{fig:lineer}) as a function of time, is plotted. The results agree qualitatively with the FP results \cite{footnote}.

We have used compatible values for corresponding parameters in the FP analysis: The exact diffusion constant $\Delta=2\alpha(1-\alpha)/(1-2\alpha)$ of the system \cite{diffconst} yields the value $\Delta=0.225$ for $\alpha=\beta=0.1$. We have also observed atypical behaviour within two lattice sites of the boundaries in the MC data, which were excluded from plots in Fig. \ref{ikili_mc}. This corresponds to a boundary smoothness of $0.08$, which too was used in the FP analysis.

Fundamental components of the system's response were calculated using the time-dependent probability function $\rho(n,t)$ for the occupation of $n$ sites at time $t$:
\begin{eqnarray}
C &=& \frac{1}{\tau} \sum_0^{\tau}\overline{n}(t) \mbox{cos}(2\pi t/\tau)  \nonumber \\
S &=& \frac{1}{\tau} \sum_0^{\tau}\overline{n}(t) \mbox{sin}(2\pi t/\tau)
\label{disc}
\end{eqnarray}
where
\begin{equation}
\overline{n}(t)=\sum_n \rho(n,t)n.
\end{equation} 
\begin{figure}       
  \begin{center}
     \includegraphics[width=8.6cm]{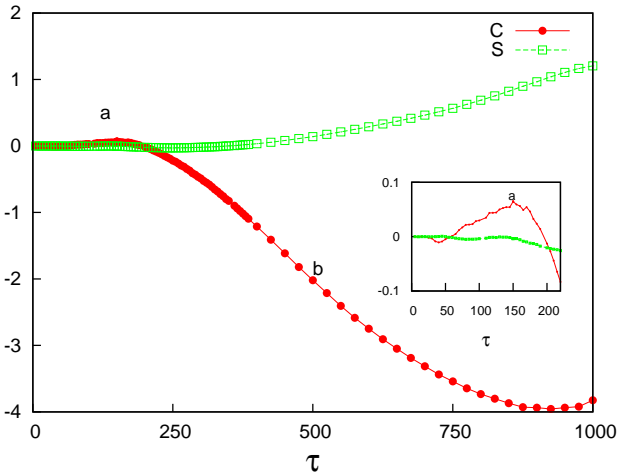}
   \end{center}
  \caption{\label{fig:mc_response} Fundamental components ($C$ and $S$ in Eqn.~\ref{disc})  with respect to different period values. Ten maximum and minimum values of particle number are excluded from the calculation. Extrema in the plot, indicate resonance like response of the system. Points $\mathbf{a}$ and $\mathbf{b}$ correspond to the density distributions displayed in FIG.~\ref{ikili_mc}. Inset shows that for smaller values of period, the interesting sinusoidal behavior is present.}
\end{figure}

Oscillations in these components are shown in Fig.~\ref{fig:mc_response}. These  oscillations are not as prominent as those obtained in the FP analysis, and displayed in Fig.~\ref{fig:ikili_fp}. We attribute this to the decay of the magnitude of the effective force away from the boundaries, damping the effect at higher frequencies.  
Time dependence of the boundary rates cause sufficient fluctuations in the TASEP system to cause the appearance of the density distributions that are quite different from those which result from constant boundary conditions. The details of these effects in the TASEP model will be reported separately. Here we have only discussed results relevant to the diffusive motion of the shock profile in this system.
\section{Conclusions}
We have demonstrated that the response of an over-damped system to a retarded oscillatory force from the boundaries leads to resonant effects in the oscillation amplitudes of statistical quantities such as the average position. These are apparent in numerical solutions of the associated FP equation as well as the Monte Carlo analysis of the motion of shock fronts in a TASEP system. 
The basis of this phenomenon is the matching of a number of density oscillations in space within the system size. The density oscillations are driven by a force field whose amplitude is space dependent, but does not have a frequency dependence. The frequency dependence in the response is the result of the matching stated above. 
\section*{ACKNOWLEDGEMENTS}
The authors acknowledge support from Turkish Academy of Sciences. The authors would also like to thank Roland Netz for useful discussions and thank to M. Ozgur Oktel for a critical reading of the manuscript.


\begin{thebibliography}{99}
\bibitem{footnote} In Fig.~\ref{ikili_mc}, the time average of the density was subtracted from $\rho (n,t)$ to magnify its time-dependece. Note that the scales in Fig.~\ref{ikili_mc} are with respect to the variable $t$, while those in Fig.~\ref{fig:ikili_fp} are with respect to $\theta$, which contains a phase shift.
\end{thebibliography}
\end{document}